\documentstyle[sprocl]{article}

\def\Journal#1#2#3#4{{#1} {\bf #2}, #3 (#4)}

\def\AAP{\em Astronomy and Astrophysics}
\def\APJ{\em The Astrophysical Journal}

\def\ARA{\em Annual Review of Astronomy and Astrophysics}

\def\MNR{\em Monthly Notices of the Royal Astronomical Society}
\def\PSP{\em Publication of the Astronomical Society of the Pacific}

\def\be{\begin{equation}}
\def\ee{\end{equation}}
\def\bea{\begin{eqnarray}}
\def\eea{\end{eqnarray}}

\begin{document}

\title{SEARCH FOR ULTRALUMINOUS X--RAY SOURCES\\IN NEARBY GALAXIES}

\author{L. Foschini$^{(1)}$\footnote{Email: \texttt{foschini@bo.iasf.cnr.it}}, G. Di Cocco$^{(1)}$,
L. C. Ho$^{(2)}$, L. Bassani$^{(1)}$, M. Cappi$^{(1)}$,\\ M. Dadina$^{(1)}$, F. Gianotti$^{(1)}$,
G. Malaguti$^{(1)}$, F. Panessa$^{(1)}$, E. Piconcelli$^{(1)}$,\\ J.B. Stephen$^{(1)}$, M. Trifoglio$^{(1)}$}

\address{$^{(1)}$Istituto di Astrofisica Spaziale e Fisica Cosmica (IASF) del CNR\\Sezione di Bologna (formerly iTeSRE), Bologna (Italy)}
\address{$^{(2)}$The Observatories of the Carnegie Institution of Washington -- Pasadena (USA)}


\maketitle\abstracts{Little is presently known about the nature of ultraluminous
X--ray sources (ULX). Different hypotheses have been
proposed to explain their properties: intermediate--mass BHs, Kerr BHs, young SNR, or
background AGN. Some of the current problems and open questions in this research field
are here reviewed.}

\section{Introduction}
The study of the X--ray emission from discrete sources in nearby galaxies began with the
Einstein satellite~\cite{GF1}. Soon after, ROSAT contributed with other important steps
and drew the attention to those pointlike X--ray sources with high luminosities
($10^{39}-10^{40}$~erg/s)~\cite{RE,CO}, while ASCA observations produced the first
spectra and lightcurves~\cite{MA}, triggering a number of hypotheses about the nature of these ultraluminous
X--ray sources (ULX). The most intriguing hypothesis is that these ULX are powered by
accretion around a black hole with masses up to $10^2-10^4$ solar masses.
Chandra, with its sub--arcsecond resolution produced a great
advancement in this study: in the observation of the Antennae galaxies, 14 ULX
were found~\cite{GF2}. Among these 14 ULX a few could well be background AGN, but certainly
not all. One of these ULX has $L_{X}\approx 10^{40}$~erg/s, similar to low luminosity AGN.
It is however located far from the centre of the host galaxy, posing several problems about
the evolution of the off--nuclear black holes in nearby galaxies.
During the last few years, with the advent of Chandra and XMM--Newton, a lot of efforts have been
done to observe and to understand these sources and how they are linked to the host galaxy.
Here we review some of the most important problems in this type of research field.

\section{Optical identification}
The search for the optical counterpart of ULX is now one of the most important
issues to understand the nature of the ULX and to ``clean'' the surveys from
background objects. One case experienced by our team can be considered paradigmatic.
During a survey searching for ULX in nearby galaxies with XMM--Newton~\cite{LF1},
we discovered one ULX in NGC4698 with $L_{X}=3\times 10^{39}$~erg/s. From online archives
we found a clear counterpart at radio (VLA 6 and 20 cm) and at optical wavelengths (DSS, HST).
Despite the variety of data none was decisive to discriminate between an ULX or a background object.
Only after we observed it with ESO--VLT, we were able to measure the redshift of $z=0.43$
(to be compared with the $z=0.0033$ of NGC4698) and it became clear that the
detected source is a background BL Lac~\cite{LF2}.

For another ULX candidate in our survey (NGC4565--ULX4), we have found two
possible counterparts in online archives: a globular cluster and a planetary
nebula, located at $8.6''$ and $8.4''$, respectively, from the XMM--Newton
position. However, Wu et al.~\cite{WU}, after a deeper analysis of HST archival data,
have found the most probable optical counterpart ($B\approx 25.1$) at less than $0.5''$
from the Chandra position and $1.8''$ from the XMM--Newton position.
They suggest that the X--ray source is an accreting black hole located in a
globular cluster at the outer edge of NGC4565.

Other possible counterparts have been found for NGC5204--X1 (HMXRB system with O star plus BH or
neutron star~\cite{RO}) and for a few ULX in NGC1399 (black holes in globular clusters\cite{LA}).
For other ULX the nearby environment has been studied, showing that these objects are often associated
with HII regions or planetary nebulae~\cite{LF1,PM,WA}. In conclusion, the search for optical identification
of ULX sources is not yet conclusive and it remains the most important question in this field.

\section{Criteria for selection and the contamination problem}
Given the present scarce knowledge about ULX, the criteria of definition and selection
are still an open question as well. Although there is general agreement to search for sources with super--Eddington
luminosities (by assuming the same distance of the supposed host galaxy), the threshold value is still matter of debate.
The name itself is still matter of discussion: ULX (ultraluminous X--ray sources or ultraluminous compact X--ray sources)~\cite{MA},
SLS (superluminous X--ray sources)~\cite{ROW}, SES (super--Eddington sources)~\cite{RO},
IXO (intermediate luminosity X--ray objects~\cite{CP} are all acronyms found in literature, but they refer to
slightly different definition criteria. Most authors suggest that the threshold luminosity should be $10^{39}$~erg/s,
because it allows to select the most theoretically challenging to explain sources (e.g. Roberts, personal
communication). However, it should be noted that there are some
known pulsars able to reach luminosities up to $10^{39}$~erg/s, because of accretion via polar cap~\cite{SMC}.
In our survey~\cite{LF1}, we used threshold value of $2\times 10^{38}$~erg/s, a value that may appear to be
too low for ``ultraluminous'' sources (ultraluminous with respect to what?), but has the advantage of being driven
by physical arguments, rather than by observational criteria. Indeed, it is the Eddington limit for
a $1.4M_{\odot}$ object, corresponding to the limit of Chandrasekhar for a white dwarf.
For a neutron star a similar limit can be set to $3M_{\odot}$, so that an object with a greater mass/luminosity can
be considered a black hole candidate.

It is however known that several factors are important in the calculation of the object mass from its luminosity:
the mode of accretion, the presence or not of a strong magnetic field, flare activity, the Eddington ratio, and so on.
For example, if one considers an Eddington ratio of $0.01-0.1$, that is a typical for black hole candidates in our
Galaxy~\cite{NO}, and that the X--ray luminosity is about $30\%$ of the bolometric luminosity,
the inferred mass from the X--ray luminosity of $2\times 10^{38}$~erg/s could be of the order of $26-260M_{\odot}$.
Therefore, although objects with X--ray luminosity greater than $10^{39}$~erg/s are surely the most intriguing one,
we stress that the search for ULX (or any other acronym you prefer) is still at an early phase and
it is meaningful to avoid thight boundaries and keep less restrictive selection rules.

If this philosophy can be true for luminosity threshold, it can be misleading in the definition
of the research area. This is generally considered to be the $D_{25}$ ellipse of the host galaxy.
However, some authors suggested a search radius of $2R_{25}$, because with smaller radii, some
famous ULX (e.g. NGC1313--X2) are excluded (e.g. Colbert, personal communication; see also Colbert and Ptak~\cite{CP}).
Whatever the selection area adopted, this has important consequences in the probability of being contaminated
by background objects.

The available literature generally provides the $\log N - \log S$ in different wavelengths~\cite{HAS,HS}.
Therefore, we can calculate the expected background sources at certain fluxes in a certain area.
Adopting the $\log N - \log S$ of XMM--Newton studies of the Lockman Hole~\cite{HAS},
one expects $0.06$ background objects per arcmin$^2$ with flux $10^{-14}$~erg~cm$^{-2}$~s$^{-1}$
in the $2-10$~keV energy band, while the number is even lower ($0.03$) in the $0.5-2$~keV energy band (same flux).
These flux limits are commonly reached by both Chandra and XMM--Newton with exposures of the order of
$10-100$~ks.

On the other hand, by adopting the space distribution of quasars in the blue
band~\cite{HS}, one expects $0.02$ quasar per arcmin$^2$ with $B>22$.

\section{Final remarks}
We have exposed here a few important open questions about the search for off--nuclear
pointlike luminous X--ray sources. We note that, in addition to the questions on the nature of these sources,
the definition of selection rules is still matter of debate.
As far as surveys are concerned, we suggest to keep lower thresholds in luminosity to increase the statistics
for variability, spectroscopic, and optical identification studies, i.e. to avoid
to cut away some interesting sources.


\section*{References}
\begin{thebibliography}{99}
\bibitem{GF1}G. Fabbiano, \Journal{\ARA}{27}{87}{1989}.
\bibitem{RE}A.M. Read et al., \Journal{\MNR}{286}{626}{1997}.
\bibitem{CO}E.J.M. Colbert and R.F. Mushotzky, \Journal{\APJ}{519}{89}{1999}.
\bibitem{MA}K. Makishima et al., \Journal{\APJ}{535}{632}{2000}.
\bibitem{GF2}G. Fabbiano et al., \Journal{\APJ}{554}{1035}{2001}.
\bibitem{LF1} L. Foschini et al., \Journal{\AAP}{392}{817}{2002}.
\bibitem{LF2} L. Foschini et al.,\emph{Astronomy \& Astrophysics}, (2002), submitted.
\bibitem{RO}T.P. Roberts et al., \Journal{\MNR}{325}{L7}{2001}.
\bibitem{WU}H. Wu et al., \Journal{\APJ}{576}{738}{2002}.
\bibitem{LA}L. Angelini et al., \Journal{\APJ}{557}{L35}{2001}.
\bibitem{PM}M.W. Pakull and L. Mirioni, \texttt{astro-ph/0202488}.
\bibitem{WA}Q.D. Wang, \Journal{\MNR}{332}{764}{2002}.
\bibitem{ROW}T.P. Roberts and R.S. Warwick, \Journal{\MNR}{315}{98}{2000}.
\bibitem{CP}E.J.M. Colbert and A. Ptak, \texttt{astro-ph/0204002}.
\bibitem{SMC}X.D. Li and E.P.J. van den Heuvel, \Journal{\AAP}{321}{L25}{1997}.
\bibitem{NO}M.A. Nowak, \Journal{\PSP}{107}{1207}{1995}.
\bibitem{HAS}G. Hasinger, \Journal{\AAP}{365}{L45}{2002}.
\bibitem{HS}F.D.A. Hartwick and D. Schade, \Journal{\ARA}{28}{437}{1990}.


\end{thebibliography}
\end{document}